\documentclass[12pt,preprint]{elsarticle}
\usepackage{graphicx}
\usepackage{amssymb}
\usepackage{amsthm}
\usepackage[all,cmtip]{xy}
\theoremstyle{plain}

\usepackage{relsize}
\begin{document}
\title{A note on relativistic rocketry 
}
\hyphenation{pos-i-tron}
\hyphenation{there-by}
\hyphenation{mat-ter-an-ti-mat-ter}
\author{Shawn Westmoreland}
\ead{westmore@math.ksu.edu}
\address{Department of Mathematics,
Kansas State University, Manhattan, Kansas, 66506, USA}
\date{October 10, 2009}
\begin{abstract}In the context of special relativity, we discuss the specific impulse of a rocket whose exhaust jet consists of massive and/or massless particles. This work generalizes  previous  results and  corrects some  errors  of  a recently published   paper by U. Walter. (The errors stem from the omission of a Lorentz factor.) We also give suggestions about how gamma ray energy could be utilized for propulsion. 
\end{abstract}
\maketitle

\section{Introduction\protect\footnote{Symbols are given in  the Appendix.}}
In a recently published paper, U. Walter  \cite{Walter} considered the problem of deriving an expression for the specific impulse of a  relativistic rocket which utilizes massless and/or massive particles in its exhaust jet. This is an exercise in Special Relativity which does not seem to have been remarked upon before Walter's paper, but unfortunately  the solution given by Walter is erroneous due to the omission of  a Lorentz factor at a crucial step. 

The main purpose of this paper is to fix Walter's solution and to expound on some consequences this has. In particular, we find that the antimatter rocket   described in Walter's paper can  achieve a specific impulse of about $0.58c$. This  is much higher than Walter's  figure  of $0.21c$, and it agrees with some calculations previously published by Vulpetti \cite{Vulpetti}. 

In addition, we give an example  in which we calculate the specific impulse for a particular design of antimatter rocket  that  utilizes both massive and massless particles for propulsion. The  example  that Walter  discussed utilized only certain massive reaction products from hydrogen-antihydrogen annihilations. The gamma rays (massless particles) are wasted in his scenario.   We consider a hypothetical modification to Walter's rocket design in which some of the gamma ray energy is utilized.

\section{Specific impulse} Consider a rocket of mass $M$ accelerating itself along a straight line through resistance-free flat space.  (``Mass" always means ``rest mass" in this paper.) Choose an inertial frame $\mathcal{F}$ that is instantaneously at rest with respect to the rocket. During an infinitesimal tick  $d\tau$ of proper time in $\mathcal{F}$ (which we regard as equivalent to an infinitesimal interval of proper time aboard the rocket), the rocket changes its momentum by an amount $Md\sigma$, where $d\sigma$ denotes the infinitesimal change in the speed of the rocket with respect to $\mathcal{F}$. (Hence $d\sigma$ is the change in the ``proper speed" of the rocket.) 

\begin{figure}[h!]
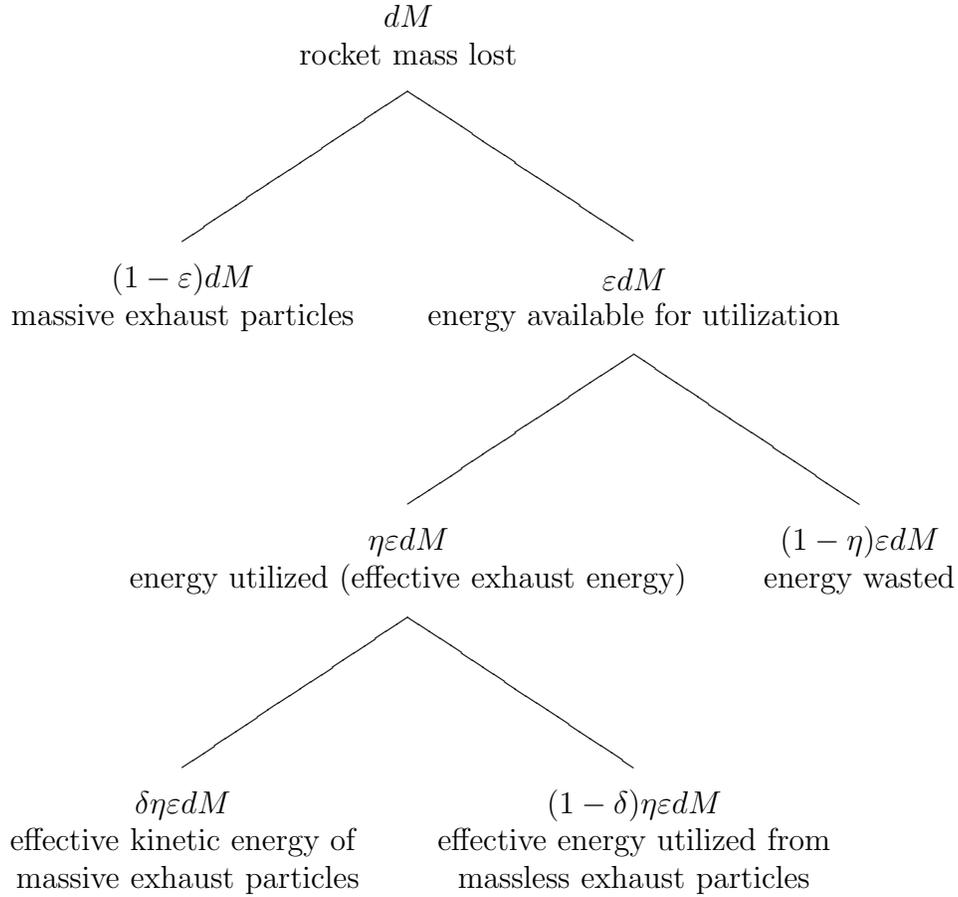

\centering
\[
\xy
(0,10)*{dM };
(0,5)*{\textrm{rocket mass lost}};
(0,0)*{}="dM"; 
(-30,-20)*{}="(1-e)dM"; (30,-20)*{}="edM"; 
(-30,-25)*{(1-\varepsilon)dM }; 
(-30,-30)*{\textrm{massive exhaust  particles}};
(30,-25)*{\varepsilon dM };
                                    (30,-35)*{}="edMA"; 
                                    (30,-30)*{\textrm{energy available for  utilization}};
(0,-55)*{}="nedM";
(0,-60)*{\eta\varepsilon dM };
(0,-65)*{\textrm{energy utilized (effective exhaust energy)}};
(0,-70)*{}="nedMA";
(60,-55)*{}="(1-n)edM";
(60,-60)*{(1-\eta)\varepsilon dM};
(60,-65)*{\textrm{energy wasted}};
(-30,-90)*{}="dnedM";
(-30,-95)*{\delta\eta\varepsilon dM };
(-30,-100)*{\textrm{effective kinetic energy of}};
(-30,-105)*{\textrm{ massive exhaust particles}};
(30,-90)*{}="(1-d)nedM";
(30,-95)*{(1-\delta)\eta\varepsilon dM };
(30,-100)*{\textrm{effective energy utilized from}};
(30,-105)*{\textrm{massless exhaust particles}};
"dM";"edM"**\dir{-};
"dM";"(1-e)dM"**\dir{-};
"edMA";"nedM"**\dir{-};
"edMA";"(1-n)edM"**\dir{-};
"nedMA";"dnedM"**\dir{-};
"nedMA";"(1-d)nedM"**\dir{-};
\endxy
\]
\caption{Energy scheme accounting for the total loss of rocket mass $dM$ and its relationship to energized exhaust during an infinitesimal interval of proper time aboard the rocket. (This diagram omits factors of $c^2$.)} 
\label{fig1}
\end{figure}

By the conservation of linear momentum, the change in the momentum of the rocket must be compensated by the ejection of propellant. In order for propellant to be ejected, the mass of the rocket must decrease. Let $dM$ denote the amount of mass lost by the rocket during the infinitesimal time interval $d\tau$. The loss in mass $dM$ is accounted for in terms of the rest masses of any massive exhaust particles together with their kinetic energies, as well massless exhaust particles and waste (refer to Figure \ref{fig1}). 

We denote by $\varepsilon$ the fractional amount of mass that is lost due to the release of energy into space. That is, the amount of energy available for propulsion (during an infinitesimal proper time interval $d\tau$) is $\varepsilon c^2 dM$. The total amount of mass lost due to the release of massive exhaust particles is then $(1-\varepsilon)  dM$. 

Some of the energy available for propulsion is likely to be wasted. Denote by $\eta$  the fractional amount of available energy that actually gets utilized for propulsion. In other words, the energy utilized by the propulsion system (during an infinitesimal interval $d\tau$ of proper time) is effectively  $\eta\varepsilon c^2 dM$ and the amount of energy wasted  is $(1-\eta)\varepsilon c^2 dM$. The wasted energy can account for, among other possibilities, a loss in efficiency resulting from an exhaust jet that forms a wide-angled cone instead of a well-collimated  beam. 

Denote by $\delta$ the  fractional amount of utilized energy that goes into the effective kinetic energy of the massive exhaust particles. That is, the massive exhaust particles have an effective kinetic energy of $\delta \eta \varepsilon c^2 dM$ and the  massless exhaust particles have an effective energy of $(1-\delta)\eta\varepsilon c^2 dM$.

Let $u$ denote the effective speed of the massive exhaust particles as they are expelled from the rocket  during the infinitesimal proper time interval $d\tau$. (The meaning of ``effective speed" is that the  massive exhaust particles affect the momentum of the rocket is as if they were all collected together into a single particle of mass $(1-\varepsilon)dM$  and thrown out of the rocket at a relative speed $u$ in the exactly backward direction.) The conservation of linear momentum gives:
\begin{eqnarray}\label{1}
Md\sigma = -\frac{(1-\varepsilon)udM}{\sqrt{1-\frac{u^2}{c^2}}} - (1-\delta)\eta\varepsilon c dM.
\end{eqnarray} The minus signs arise because  $dM$ represents a loss in rocket mass.

Note that Equation (\ref{1}) corresponds to the unnumbered equation appearing before Equation (9) in Reference \cite{Walter}, but the referenced paper omitted a Lorentz factor of $1/\sqrt{1-u^2/c^2}$ in the first term. This was the source of a significant error. 

The total effective kinetic energy of the massive exhaust particles is $\delta \eta \varepsilon c^2 dM$. Using the relativistic formula for kinetic energy, we get that:
\begin{eqnarray}\label{2}
\delta\eta\varepsilon c^2 dM = \left(\frac{1}{\sqrt{1-\frac{u^2}{c^2}}}-1\right)(1-\varepsilon)c^2 dM.
\end{eqnarray} 
Equation (\ref{2}) can be solved for $1/\sqrt{1-u^2/c^2}$ to give:
\begin{eqnarray}\label{3}
\frac{1}{\sqrt{1-\frac{u^2}{c^2}}} = \frac{1-\varepsilon(1-\delta\eta)}{1-\varepsilon}.
\end{eqnarray}
Solving Equation (\ref{3}) for $u$:
\begin{eqnarray}\label{4}
u=c\sqrt{1-\left(\frac{1-\varepsilon}{1-\varepsilon(1-\delta\eta)}\right)^2}.
\end{eqnarray}
Substituting Equations (\ref{3}) and (\ref{4}) into Equation (\ref{1}), and simplifying:
\begin{eqnarray}\label{5}
Md\sigma = -c\left(\sqrt{\delta\eta\varepsilon(2-2\varepsilon + \delta\eta\varepsilon)} + (1-\delta)\eta\varepsilon \right)dM.
\end{eqnarray}
Specific impulse $w$ is defined such that (e.g., \cite{Sutton} pp. 28 - 29):
\begin{eqnarray}\label{6}
M\frac{d\sigma}{d\tau}=-w\frac{dM}{d\tau}.
\end{eqnarray} (We warn the reader that ``specific impulse" is traditionally defined as $w$ divided by the acceleration of  gravity due to  Earth at sea-level.)

From Equations (\ref{5}) and (\ref{6}), we obtain the following expression for $w/c$:
\begin{eqnarray}\label{7}
\frac{w}{c} = \sqrt{\delta\eta\varepsilon(2-2\varepsilon + \delta\eta\varepsilon)} + (1-\delta)\eta\varepsilon. 
\end{eqnarray}

Equation (\ref{7}) fixes the error in Equation (16) of Reference \cite{Walter} and  generalizes existing results in the literature. In the case where $\eta=\delta=1$, Equation (\ref{7}) reduces to $w/c=\sqrt{2\varepsilon-\varepsilon^2}$, which agrees with the corresponding result derived by S\"anger in Section 2  of his 1953 paper ``Zur Theorie der Photonraketen"    \cite{Sanger}. The case where $\eta=1$ and $\delta=0$ can be interpreted as  a  photon rocket that simply jettisons spent fuel at zero relative speed; and here Equation (\ref{7}) reduces to $w/c=\varepsilon$,  in agreement with  results derived in Section 3c of  Reference \cite{Sanger}.

The specific impulse of a practical interstellar rocket would need to be  a significant fraction of the speed of light. 
In order to achieve this, one must  convert mass into energy with nearly perfect efficiency.  It is commonly assumed that the only known  way of doing this  involves the annihilation of matter with antimatter. However, another possibility involves the quantum mechanical evaporation of a black hole. Crane \emph{et al.} \cite{CW} argue that a micro-black hole with a Schwarzschild radius on the order of a few attometers would be an excellent  power source for an interstellar rocket. Moreover, Crane argues that it would be easier to make   black holes of the requisite size than it would be to make   large quantities of antimatter  needed to drive an interstellar starship. Also, black holes would be safer and easier to use than antimatter. For details see Reference \cite{CW}. In the remainder of this paper, we will use available literature on the more familiar concept of the antimatter rocket as an illustration of the use of Equation (\ref{7}).

\section{An application to antimatter rockets}

The purpose of this section is to use Equation (\ref{7}) to reassess the maximum  specific impulse achievable by the  antimatter rocket studied in Reference \cite{Walter}. This rocket annihilates hydrogen with antihydrogen and uses electromagnetic fields to collimate charged reaction products into an exhaust beam. Gamma rays, which are also produced in the annihilation, escape into space and their energy is not utilized. (In Section 4, we will consider the possibility of utilizing gamma ray energy.)

Table \ref{table1} describes  what happens, on the average, when an atom of hydrogen annihilates  with an atom of antihydrogen at rest. The electron from the hydrogen atom annihilates with the positron from the antihydrogen atom and  a pair of gamma rays results. The proton from the hydrogen atom annihilates with the antiproton from the antihydrogen atom and the initial result is, on the average, about two neutral pions $\pi^0$ and three charged pions ($\pi^+$ and $\pi^-$ particles) \cite{Frisbee}. The neutral pion is extremely short-lived and only traverses a microscopic distance before giving rise to its decay products, 
which are usually (i.e., $98.798\pm0.032\%$ of the time \cite{PDG}) two gamma rays. On the other hand, the charged pions travel a good macroscopic distance (on the order of a couple tens of  meters) before giving rise to their decay products, 
which are usually (i.e., $99.98770\pm0.00004\%$ of the time \cite{PDG}) just a muon $\mu^+$ (or antimuon $\mu^-$) together with a muon neutrino $\nu_\mu$ (or antimuon neutrino $\bar{\nu}_{\mu}$). The muons and antimuons  travel a distance on the order of a kilometer before decaying (into electrons, positrons and neutrinos).

\begin{table}[h]
\begin{center}
\begin{tabular}{l|c|c}
species& rest mass (MeV) & kinetic energy (MeV)\\
\hline
\hline
inital reactants:& &\\
$p^+$ & 938.3& 0\\
$e^-$ & 0.5& 0\\
$p^-$ & 938.3& 0\\
$e^+$ & 0.5& 0\\
\hline
initial products:& &\\
$2.0\pi^0$ & 269.9 & 439.1\\
$1.5\pi^+$ & 209.4 & 374.3\\
$1.5\pi^-$ & 209.4 & 374.3\\
$2\gamma$ (from $e^-+e^+$)& 0 & 1.0\\
\hline
decay products:& & \\
$4\gamma$ (from $2\pi^0$)& 0 & 709.1\\
$1.5 \mu^+ $ (from $1.5\pi^+$)&158.5&288.5\\
$1.5\nu_\mu$ (from $1.5\pi^+$)& 0& 136.8\\
$1.5\mu^-$ (from $1.5\pi^-$)& 158.5&288.5\\
$1.5\bar{\nu}_\mu$ (from $1.5 \pi^-$)& 0 & 136.8\\
\end{tabular}
\caption{Reaction products arising from the annihilation of hydrogen with  antihydrogen at rest (data taken from Reference \cite{Frisbee}).}\label{table1}
\end{center}
\end{table} 

The antimatter rocket design in Reference \cite{Walter} achieves its thrust by  collimating (via  electromagnetic fields)  the charged pion products  into an exhaust jet. The gamma rays simply escape into space as waste. A negligible amount of reaction products are absorbed by  the spacecraft. This particular antimatter rocket design is often discussed  elsewhere in the literature (see, for example, Frisbee \cite{Frisbee} and references therein). 

Referring to Table \ref{table1}, and  assuming that the charged pion products are collimated with perfect efficiency, we get that:
\begin{eqnarray}\label{1-e}
1- \varepsilon&=&\frac{\textrm{massive exhaust particles utilized}}{\textrm{rocket mass lost}}\nonumber\\
&=& \frac{418.8}{1877.6},
\end{eqnarray} which yields (in agreement with Reference \cite{Walter}):
\begin{eqnarray}
\varepsilon = 0.7769.
\end{eqnarray} Moreover, since only the kinetic energy of the charged pion products are utilized for propulsion, we get that:
\begin{eqnarray}
\eta\varepsilon&=&\frac{\textrm{energy utilized}}{\textrm{rocket mass lost}}\nonumber\\
&=&\frac{748.6}{1877.6},
\end{eqnarray}
yielding (also in agreement with Reference \cite{Walter}):
\begin{eqnarray}
\eta = 0.5132.
\end{eqnarray} Furthermore, since this example utilizes only massive  particles as exhaust, we have that $\delta = 1$. 

Plugging  the values $\varepsilon = 0.7769$, $\eta=0.5132$, and $\delta =1$ into Equation (\ref{7}) gives:
\begin{eqnarray}\label{sipion}
\frac{w}{c} = 0.5804.
\end{eqnarray} Thereby we find that the ideal specific impulse of the rocket is $0.5804c$. This is significantly higher than the value of $0.2082c$ obtained in Reference \cite{Walter}.

Vulpetti \cite{Vulpetti} supports our $0.5804c$ result. Indeed, Equation (6) of Vulpetti's paper implicitly defines an expression for specific impulse (Vulpetti assumes a purely massive-exhaust drive) which  is equivalent to our Equation (\ref{7}) if $\delta=1$. The fact that Vulpetti's equation is implicit in Equation (\ref{7}) may not be  obvious at first glance because Vulpetti's equations are  expressed in terms of   variables that are quite different from ours. 

\section{On utilizing  gamma ray energy for propulsion}

The propulsion design discussed in Section 3 utilizes, at best, only  $\varepsilon\eta = 39.87\%$ of the annihilation energy  as exhaust energy (cf. Morgan \cite{Morgan}, p. 536). A large amount of  energy is uselessly carried off into space by gamma rays.

Let us consider possible ways in which the  ``pion drive" of Section 3   might be modified so that  gamma ray energy can be utilized for propulsion.  

 S\"anger famously proposed that one would need to create an extremely dense ``pure electron gas" in order to  reflect    gamma rays efficiently  \cite{Sanger2}. A parabolic reflector of this kind, with the annihilation point at its focus, would steer gamma rays into a well-collimated exhaust beam. However, the  feasibility of this proposal is unclear (see, e.g. Forward  \cite{Forward2}).

Vulpetti has proposed a method of utilizing gamma ray energy  by taking advantage of  pair production phenomena (see, e.g., Reference \cite{Vulpetti} or \cite{Vulpetti83}). The gamma rays produced by proton-antiproton annihilations are of such a high energy that, by interacting with the electric field of a  nucleus, 
 they can be converted into real electron-positron pairs. Since they are charged particles, these electrons and positrons can  be collimated by way of electromagnetic fields. 

Another alternative is to use  a gamma absorbing shield. The shield will  reradiate, in all directions,  the energy that it absorbs. The reradiated photons will tend to have optical or nearly optical wavelengths and so  can be easily  collimated. This was suggested to the author by Louis Crane. A similar concept has been discussed by Smith, \emph{et al.} \cite{Smith}, and by S\"anger (\cite{Sanger}, p. 224, second paragraph).

If we denote by $\alpha$ the fractional amount of  gamma ray energy that can be utilized in a suitably modified pion drive, then (assuming that no  reaction products are \emph{permanently} absorbed by  the spacecraft):
\begin{eqnarray}
(1-\delta)\eta\varepsilon &=& \frac{\textrm{massless exhaust particles utilized}}{\textrm{rocket mass lost}}\nonumber\\
&=& \frac{710.1\alpha}{1877.6}.
\end{eqnarray} Assuming that the pions are collimated with perfect efficiency, we have that:
\begin{eqnarray}
\delta\eta\varepsilon = \frac{748.6}{1877.6}.
\end{eqnarray} As before, we still have $1- \varepsilon = 418.8/1877.6$ (Equation \ref{1-e}). 

Plugging these into Equation (\ref{7}) gives:
\begin{eqnarray}
\frac{w}{c} = 0.5804 + 0.3782\alpha.
\end{eqnarray}

If all of the pions and gammas are utilized, then the specific impulse can be nearly  $0.96c$. If half of the gammas are utilized, then the specific impulse can be nearly $0.77c$.

\section{Conclusions and closing remarks}
In this paper, we deduced an equation, Equation (\ref{7}), which expresses (in terms of parameters $\varepsilon, \eta$ and $\delta$) the specific impulse of a rocket which utilizes massive and/or massless particles as exhaust. The analysis was done in the context of Special Relativity. We  solved a problem which was purportedly solved in a previous paper entitled ``Relativistic rocket and space flight" by U. Walter \cite{Walter}, but  Walter unfortunately omitted a Lorentz factor  which lead him to obtain   erroneous results. 

When  Equation (\ref{7}) is applied  to the case of a particular example, as  in Section 3, we find that the corrections it makes to Walter's calculations are very significant. Walter considered the problem of calculating the best specific impulse that could  theoretically  be achieved by an antimatter pion drive. He calculated a specific impulse of about $0.21c$, whereas we calculated a specific impulse of about $0.58c$.   Vulpetti \cite{Vulpetti} agrees with our  $0.58c$ result.   

In Section 4, we considered the possibility that better efficiency and even higher specific impulses could be achieved  if a way to utilize gamma ray energy could be found. We point out that even if gamma ray reflectors are not feasible,  gamma ray energy might still be utilized. For example, a gamma ray absorbing shield will radiate  back into space the energy it absorbs. Moreover, the re-emitted radiation coming from the shield will be  in the form of photons at near-optical frequencies. Thereby, the re-emmitted radiation can be  collimated into an exhaust beam with relative ease. We calculated that if a pion drive were so equipped that it could effectively  utilize half of the gamma ray energy for propulsion, then it could achieve a specific impulse of up to nearly $0.77c$.

\section{Acknowledgements} I wish  to thank my doctoral advisor, Louis Crane, for his very  helpful suggestions and encouragement. 
\section*{Appendix}

\
\\
\emph{Symbols}

\
\\
$\alpha=$ fractional amount of gamma ray energy that is effectively utilized for propulsion \\
$\gamma=$  photon \\
$\delta=$  fractional amount of propulsive energy
that goes into the effective kinetic energy
of massive exhaust particles\\
$\varepsilon=$  fractional amount of lost rocket mass that   
is accounted for by mass converting into energy\\
$\eta=$ fractional amount of available energy that is utilized for propulsion\\
$\mu^+=$   antimuon \\
$\mu^-=$  muon \\
$\nu_\mu=$   muon neutrino \\
$\bar{\nu}_\mu=$   antimuon neutrino \\
$\pi^0=$  neutral pion\\
$\pi^+=$ positive pion \\
$\pi^-=$  negative pion \\
$\sigma=$ proper speed\\
$\tau=$  proper time\\
$c=$  the speed of light (exactly 299792458 meters per second \cite{PDG})  \\
$d=$ differential operator/``infinitesimal" prefix\\
$e^+=$  positron\\
$e^-=$    electron\\
$\mathcal{F}=$ inertial reference frame instantaneously at rest with respect to the rocket\\
$M=$  instantaneous mass of rocket\\
$p^+=$ proton\\
$p^-=$   antiproton\\
$u=$  effective relative speed of massive exhaust\\
particles with respect to the rocket\\ 
$w=$ specific impulse\\


\begin{thebibliography}{}
    \bibitem{Walter} U. Walter, (2006)  ``Relativistic rocket and space flight," Acta Astronautica, Vol. 59, pp. 453 - 461.
    \bibitem{Vulpetti}G. Vulpetti, (1985) ``Maximum terminal velocity of relativistic rocket," Acta Astronautica, Vol. 12, No. 2, pp. 81 - 90.
\bibitem{Sutton}G. P. Sutton, O. Biblarz, \emph{Rocket Propulsion Elements}, 7th Edition, John Wiley and Sons, New York, 2001.
\bibitem{Sanger}E. S\"anger,  (1953) ``Zur Theorie der Photonraketen," Ingenieur-Archiv, Band 21,  pp. 213 - 226 (in German).
\bibitem{CW}L. Crane and S. Westmoreland, (2009) ``Are black hole starships possible?" arXiv:0908.1803 [gr-qc].
\bibitem{Frisbee} R. H. Frisbee, (2003) ``How to build an antimatter rocket for interstellar missions," 39th AIAA/ASME/SAE/ASEE Joint Propulsion Conference and Exhibit, Huntsville, Alabama, July 20 - 23, 2003.
\bibitem{PDG}C. Amsler, \emph{et al.}, (2008) ``Review of Particle Physics," Physics Letters B, Vol. 667, 1.


\bibitem{Morgan} D. L. Morgan, (1988) in \emph{Antiproton Science and Technology}, ed. R. W. Augenstein \emph{et al.}, pp. 530 - 565.
\bibitem{Sanger2}E. S\"anger, ``Photon propulsion," in \emph{Handbook of Astronautical Engineering}, First Edition, H. H. Koelle (Ed.), McGraw-Hill, New York, 1961.
\bibitem{Forward2} R. L. Forward, (1985) ``Antiproton annihilation propulsion," J. Propulsion, Vol. 1, 370 - 374.

\bibitem{Vulpetti83}G. Vulpetti, (1983) ``A concept of low-thrust relativistic-jet-speed high-efficiency matter-antimatter annihilation thruster," International Astronautical Federation 34th Congress, Budapest, October 1983.
\bibitem{Smith}D. W. Smith, et al., (2005) ``Thermal radiation studies for an electron-positron annihilation propulsion system," 41st  AIAA/ASME/SAE/ASEE Joint Propulsion Conference, Tucson, Arizona, July 10 - 13, 2005.

  \end{thebibliography}
\end{document}


\section*{Errata}
\
\\
After some revision, my paper ``A note on relativistic rocketry" was accepted for  publication. The main results (Equations (7) and (12)) did not need to be changed, and their justifications also remained intact. Here is the bibliographical info: \emph{Acta Astronautica}, Volume 67, Issues 9-10, November-December 2010, pp. 1248 - 1251.

 \
 \\
 A  pdf file that matches the published version will, at least temporarily, be legally available for free download at my personal website:
 
\
\\
  \verb+http://www.math.ksu.edu/~westmore/articles+

\
\\ 
The article is also available at the following stable  URL:

\
\\
 \verb+http://dx.doi.org/10.1016/j.actaastro.2010.06.050+

\
\\
I would have liked to upload to arXiv a replacement that matches the published version, but this is not allowed by Elsevier's copyright policy. Since I cannot upload the corrected version of my preprint, I am uploading this ``errata" instead.  

\begin{itemize}
\item{The \emph{primary shortcoming} of the original preprint, which initially prevented its publication, had to do with the proposed method of utilizing gamma rays for propulsion (Section 4 in the original preprint). One of the reviewers of the article pointed out serious problems with that proposal and so it was omitted from the published version. The published version offers no suggestions on \emph{how}  to utilize gammas and warns the reader that this is a very difficult problem with no clear solution at the present time.

The main reason why my originally proposed  method would not work has to do with the radiator. Unless the radiator is made from some kind of extremely low-density material that absorbs gamma rays well --- and I do not know of any such material --- then the radiator will have such  a huge mass that its presence would severely hinder rather than aid the performance of the spacecraft.

The following back-of-the-envelope calculations support this assertion.\footnote{I should make it clear that these calculations do not appear in my published article. The purpose of presenting these calculations here is to explain why some parts of my original article had to be changed.}  In order for a pure photon rocket of mass $M$ to maintain a proper acceleration $a$, an exhaust power of $acM$ is required.\footnote{See, e.g., L. R. Shepherd, ``Interstellar Flight," in \emph{Realities of Space Travel: Selected Papers of the British Interplanetary Society}, ed. by L. J. Carter, McGraw-Hill, New York, 1957.} Suppose that the radiator can operate at a maximum temperature of $T$ and assume that the radiator is a blackbody built like a  sphere centered on the gamma source. It follows from the Stefan-Boltzmann law that the minimum  radius of the radiator is: $$r=\sqrt{\frac{L}{4\pi\sigma T^4}},$$ where $L$ ($=acM$) is the luminosity of the gamma source and $\sigma$ is the Stefan-Boltzmann constant. (We are ignoring the possibility that the radiator might absorb some of its own radiation.) Let $\rho$ denote the density of the radiator and let $x$ denote its thickness. Assuming $x\ll r$, the mass of our minimal radiator is approximately $M_{rad}=4\pi r^2 \rho x$. Using the above relation for $r$, together with $L=acM$ and $M_{rad} = 4\pi r^2 \rho x$, it follows that:
$$\frac{Ma}{M_{rad}} = \frac{\sigma T^4}{c\rho x}.$$ Since $M_{rad} <M$, this implies that:
$$a< \frac{\sigma T^4}{c\rho x}.$$ Although interstellar travel can get away with very low accelerations sustained over a long time, we should be disappointed if the upper bound for $a$ turns out to be extremely  small. 

Let us estimate the upper bound on $a$  corresponding to a  tungsten radiator. Tungsten is a good gamma absorber, and so we shall assume that  a tungsten object bombarded with gamma radiation behaves like a blackbody. Moreover, since tungsten has an extremely high melting point compared to other known materials, it is a natural choice for our radiator. We note that Frisbee [6] discusses some apparently plausible uses for tungsten radiators for interstellar rockets --- however  he does not make the mistake of proposing that  the thermal energy from his radiators  be utilized for propulsion. Frisbee's hypothetical radiators have a width of $0.1$ meters and are assumed to operate at a temperature of 1500 K (though we note that the melting point of tungsten is quite higher than this).  According to Wikipedia, the density of tungsten ranges from about 17.6 grams per cubic centimeter (at melting point) to about 19.25 grams per cubic centimeter (at room temperature). The melting point of tungsten, according to Wikipedia, is 3695 K. In order to get an optimistic result (which will still turn out to be discouraging), let us take $\rho = 17.6$ g/cm$^{3}$, $T=4000$ K and use a thickness of $x=0.1$ m. This gives an  upper bound for $a$  on the order of $2.8\times10^{-5}$ m/s$^{2}$. Note that with a constant proper acceleration of   this magnitude, it would take more than 800 years to cover the first light year. Moreover, after all that time, the gain in speed would be less than a quarter  of $1\%$ of the speed of light.

Based on these discouraging calculations, it seems that in order for a photon drive to run on \emph{thermalized} energy as I had previously proposed, one needs a radiator that (1) can operate at a very temperature, (2) has a low density, and (3) absorbs gamma rays sufficiently well  that its thickness can be kept small. I am not sure whether  there is any such material --- solid, liquid, gas, or plasma --- that satisfies all three of these properties. } 

\item{(This is not a correction but a remark.) In my final revision, I inserted  a short paragraph (appearing in the published version, second to last paragraph in Section 2) concerning the distinction between  ``specific impulse" and ``effective exhaust speed." I do not know if this distinction is well-known or not.\footnote{Note that Vulpetti [2] apparently  uses it since he distinguishes between ``true jet speed" ( = ``effective relative speed of
the exhaust particles," in my terminology) and ``effective exhaust speed"
(= ``specific impulse," in my terminology).} 

 The distinction that I am talking about is both conceptual and quantitative.  Simply put, Equation (1) gives an operational definition for effective exhaust speed $u$ and Equation (6) gives an operational definition for  specific impulse $w$. These two definitions are conceptually independent. Moreover, the quantitative values of $u$ and $w$, as given by Equations (4) and (7), are not equivalent. Note however (as I failed to mention in the published version) that in the ``non-relativistic limit" of Equations (1) and (6), where $\varepsilon\rightarrow 0$ and $c\rightarrow \infty$, one gets that $w=u$. }
\end{itemize}

\
\\
Shawn Westmoreland
\
\\
Kansas State University
\
\\
September 3, 2010